# Impedance-controlled Laser Dynamics in Near-Zero Index Media

Adrien Debacq,[1] Andrei V. Lavrinenko,[2] Elizaveta Semenova, [2,3] Olivier Deparis,[1] and Michaël Lobet[1,*]

[1]University of Namur, Department of Physics & Namur Institute of Structured Materials, Rue de Bruxelles 61, 5000 Namur, Belgium
[2]DTU Electro-Department of Electrical and Photonics Engineering, Technical University of Denmark, DK-2800 Kongens Lyngby, Denmark
[3]NanoPhoton - Center for Nanophotonics, Technical University of Denmark, DK-2800 Kongens Lyngby, Denmark

**Abstract** - Near-zero-index (NZI) media have been explored for controlling spontaneous emission, but their impact on stimulated emission and lasing properties remains unexplored within their different classes, i.e., epsilon-near-zero (ENZ), mu-near-zero (MNZ), or epsilon-and-mu-near-zero (EMNZ). Here, we show theoretically that lasing emission depends on the class of NZI medium and can be controlled through its impedance. The major difference, as compared to conventional laser media, arises from the incorporation of dispersive electric and magnetic properties in the derivation of the Maxwell Boch equations. In this framework, the analysis of laser rate equation reveals that the ENZ class, with high impedance, can reduce the lasing threshold through enhanced electric-field interaction. Numerical simulations further show that the NZI class also governs the spectral response: whereas ENZ and MNZ classes suffer from multimode emission, the EMNZ class preserves single-mode emission.

The electromagnetic environment plays a central role in shaping light–matter interactions. Since the pioneering work of Purcell [1], it is known that the spontaneous emission rate of an emitter can be strongly modified by its environment, either through optical confinement in cavities [2] or through the engineering of the local electromagnetic density of states in structured media [3].

In this context, near-zero-index (NZI) media have been shown to strongly modify spontaneous emission in a manner that depends on the specific NZI class and dimensionality [4–11]. NZI media are classified as epsilon-near-zero (ENZ), mu-near-zero (MNZ), or epsilon-and-mu-near-zero (EMNZ), depending on whether the relative permittivity, permeability, or both approach zero [12–16]. This classification gives rise to markedly different electromagnetic responses: ENZ and MNZ media can support slow-light propagation, while their impedances tend toward infinity and zero, respectively, for infinitely homogeneous media [4,17,18]. In contrast, both the impedance and group refractive index of EMNZ media can be tuned through their dispersive properties [19–23]. These differences strongly affect the balance between electric- and magnetic-field interactions, leading to their pronounced enhancement in ENZ and MNZ media, respectively [14,24]. Such unusual electromagnetic responses have been extensively exploited to control spontaneous emission, enabling enhanced emission rates [4–9] and long-range collective effects such as superradiance [25–29].

Despite these advantages, the effects of NZI environments on stimulated emission and laser dynamics remain largely unexplored [4,21]. For instance, it remains unclear how the individual NZI classes influence the lasing threshold, modal properties, and laser dynamics. Addressing this question requires disentangling the respective roles of the dispersive permittivity, permeability, group velocity, and wave impedance in light–matter interactions.

In this Letter, we investigate theoretically how the class of NZI media governs the threshold and spectral properties of solid-state lasers. By incorporating the electric and magnetic dispersive properties of the active medium analytically into the laser rate equations, we reveal that light–matter interactions are governed not only by the group refractive index, but also by the wave impedance. Specifically, the high impedance of ENZ media, combined with their ability to slow down light velocity, enhances the electric-field interactions and reduces the lasing threshold, whereas the low impedance of MNZ media weakens the electric-field interactions and increases the lasing threshold. Likewise, the lasing threshold in EMNZ media can be controlled through their impedance. Using finite-difference time-domain (FDTD) simulations, we further examine the spectral response. The large group indices of ENZ and MNZ media lead to multimode emission, whereas EMNZ media preserve single-mode lasing despite their near-zero refractive index. Together, these results reveal how the electromagnetic parameters of NZI media, i.e. group index and impedance, independently shape the threshold and spectral properties of lasers.

To disentangle the respective roles of the active medium permittivity, permeability, and group index in laser emission, we generalized the Maxwell–Bloch formalism from which laser rate equations are commonly derived. The active medium is modeled as an ensemble of two-level emitters described by conventional Bloch equations for the population inversion and microscopic polarization [30,31]. However, the frequency-dependent permittivity and permeability of the active medium are explicitly retained in Maxwell's equations, thereby modifying the effective coupling between the electromagnetic field and the atomic polarization.

*Contact author: michael.lobet@unamur.be

The electromagnetic field is considered inside an optical linear cavity filled with this dispersive active medium. Expanding the electromagnetic response around the laser frequency and applying the slowly varying envelope approximation [30,32] (see Appendix A), the temporal evolution of the electric field amplitude $E(t)$ of the resonant cavity mode [32] is described by

$$\frac{dE(t)}{dt} = iv_g\left(k_{p,0} - k_{p,c}\right)E(t) - \frac{\gamma_c}{2}E(t) + i\sqrt{\frac{\mu}{\varepsilon}}\frac{v_g\omega_0}{2\varepsilon_0 c}P_{at}(t) \quad (1)$$

where $v_g$ is the group velocity, $k_{p,0} = \omega_0 n(\omega_0)/c$ and $k_{p,c} = \omega_c n(\omega_c)/c$ are the wave vectors related to the propagation $\omega_0$ and cavity resonance $\omega_c$ angular frequencies, respectively, $n$ is the phase refractive index, $\gamma_c$ is the cavity loss rate, and $P_{at}$ is the atomic polarization. The first term in Eq. (1) describes the frequency detuning between the atomic transition and the cavity resonance. In conventional media, one typically assumes $n_g \simeq n$, such that $v_g \simeq c/n$ and this first term reduces to the usual form $i(\omega_0 - \omega_c)E(t)$. This approximation, however, is no longer valid in NZI media, where the group index $n_g$ can differ drastically from the phase index $n$ and the detuning must therefore retain its explicit dependence on the group velocity. In conventional laser rate equations, the light–matter coupling, i.e., the third term of Eq. (1), is described by the coefficient $v_g i\omega_0/2n\varepsilon_0 c$, where the phase refractive index $n$ appears in place of the wave impedance $Z = \sqrt{\mu/\varepsilon}$ [30,32]. By explicitly accounting for the frequency-dependent magnetic permeability, the wave impedance emerges as the parameter governing the strength of the light–matter coupling.

From these dispersive Maxwell–Bloch equations, the laser rate equations can be derived following the usual procedure for a semi-conductor laser hereafter considered [30] (see Appendix B). The resulting carrier $N$ and photon $S$ density dynamics are given by

$$\frac{dN}{dt} = \Lambda_p - Zv_g g(N)S - \gamma_{nr}N,$$

$$\frac{dS}{dt} = n\mu\beta\gamma_{sp}N + Zv_g g(N)S - \gamma_c S, \quad (2)$$

where $\Lambda_p$ is the pumping rate, $g(N) = g_0(N - N_0)$ is the material gain (in $m^{-1}$) with $g_0$ the gain constant (in $m^2$) and $N_0$ the carrier density at transparency, $\gamma_{nr}$ is the nonradiative recombination rate, $\gamma_{sp}$ and $\beta$ are the spontaneous decay rate and beta factor (i.e., the fraction of spontaneous emission coupled to the lasing cavity mode), respectively, and $\gamma_c$ is the cavity loss rate. The $\varepsilon$- and $\mu$-dependence of $\gamma_{sp}$ is taken from the dispersive derivations of Ref. [4]. The corresponding dependence of stimulated emission derived herein depends explicitly on $n_g = c/v_g$ and $Z$, in agreement with previously reported descriptions of stimulated Einstein coefficient in dispersive media, i.e., $B'_{21} = \frac{1}{n_g}\sqrt{\frac{\mu}{\varepsilon}}B_{21}$ [4,33]. The above formalism therefore provides a framework to disentangle the respective contributions of dispersive permittivity, permeability, group velocity and wave impedance to laser emission.

We first solve the laser rate equations, Eqs. (2), for a conventional InGaAsP active medium, used as reference. The material parameters are fixed to $g_0 = 2.50 \times 10^{-16}\text{cm}^2$, $N_0 = 1.00 \times 10^{18}\text{cm}^{-3}$, $\gamma_{nr}^{-1} = \gamma_{sp}^{-1} = 2.20\text{ns}$, $\beta = 1.00 \times 10^{-4}$, and $\gamma_c^{-1} = 1.60\text{ps}$ [37], with the active medium modeled as a dielectric with $n = 3.40$ and $n_g = 4.00$ [34]. We then compare this reference with hypothetical dispersive media described by a Drude-Lorentz model, while keeping all other rate-equation parameters unchanged [14]. The parameters of $\varepsilon$ and $\mu$ models are chosen to match the target values at the lasing frequency, with the group index calculated from their frequency derivatives. We first consider two limiting cases: an ENZ medium with $\varepsilon = 0.01$ and $\mu = 1.00$, and an MNZ medium with $\varepsilon = 1.00$ and $\mu = 0.01$. Although both media share the same large group index, ($n_g = 22.01$), their impedances differ substantially ($Z_{ENZ} = 10.00$ and $Z_{MNZ} = 0.10$), enabling us to isolate the role of impedance in cavity dynamics under slow-light conditions (see Table I).

Figure 1(a) shows the steady-state laser output power as a function of injection current. The ENZ medium reduces the lasing threshold by approximately a factor of two ($I_{th}$= 7.52mA) relative to the reference ($I_{th}$= 15.52mA), whereas the MNZ medium doubles it ($I_{th}$= 31.52mA). Despite their identical group indices, ENZ and MNZ therefore produce opposite shifts of the lasing threshold, indicating that slow light with the same group velocity can interact differently with matter depending on the impedance. The high impedance of the ENZ medium enhances the electric field, thereby strengthening the electric-dipole interactions with the active medium, whereas the low impedance of the MNZ medium enhances the magnetic field and thus favors the magnetic response at the expense of the electric-dipole interactions.

*Contact author: michael.lobet@unamur.be

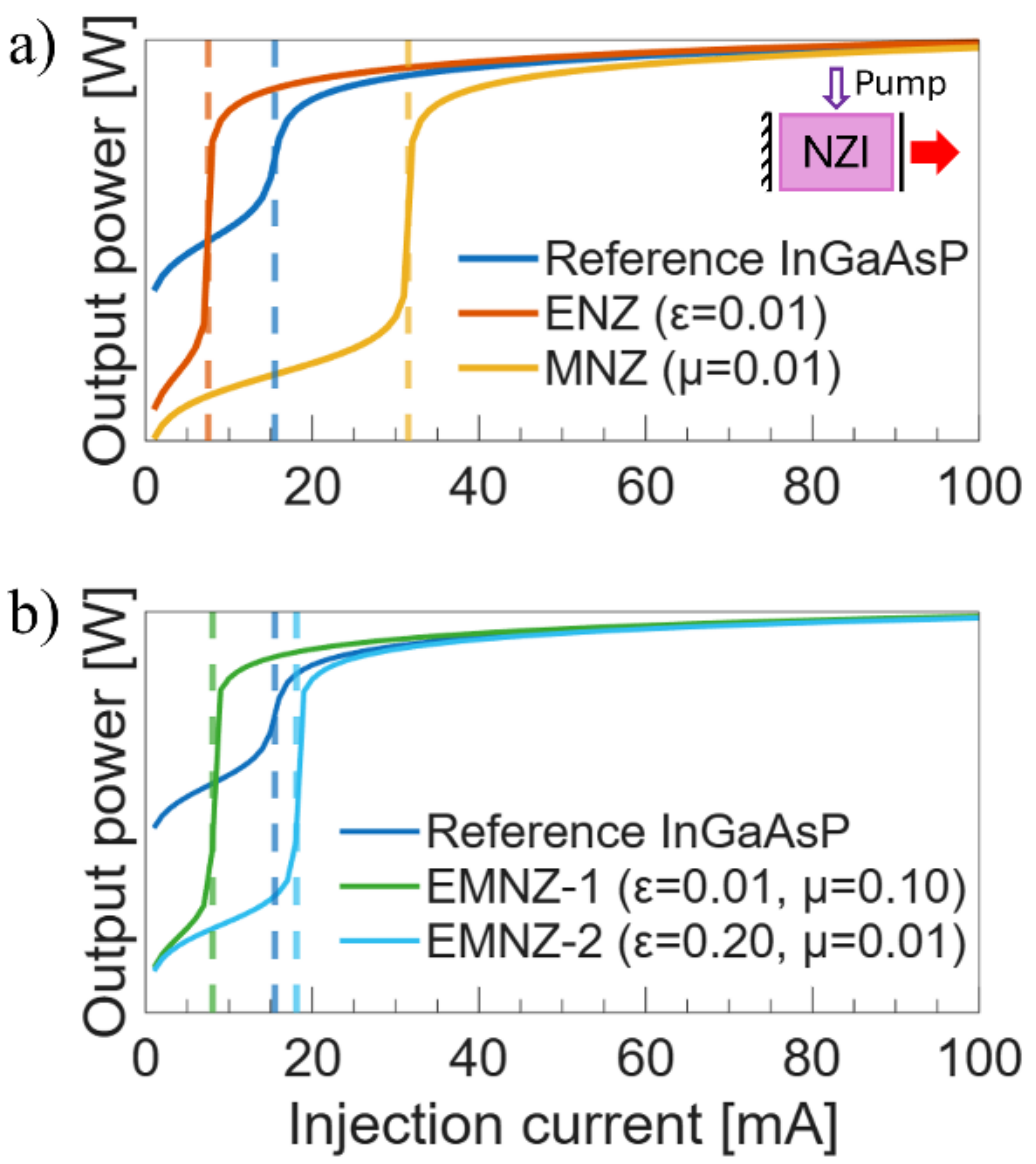

FIG 1. Light–current characteristics obtained from solving the laser rate equations for (a) the reference InGaAsP (blue), ENZ (red), and MNZ (yellow) media with a schematic of the laser linear cavity as inset, and (b) the reference InGaAsP and two EMNZ media (green and light blue). Dashed lines indicate the corresponding lasing thresholds.

The role of the impedance is further highlighted by the EMNZ class, for which the impedance depends on the relative dispersion of both $\varepsilon$ and $\mu$ [4]. We therefore consider two EMNZ media with different combinations of permittivity and permeability, EMNZ-1 with $\varepsilon = 0.01$, $\mu = 0.10$ ($Z_{EMNZ-1}$ =3.16) and EMNZ-2 with $\varepsilon = 0.20$, $\mu = 0.01$ ($Z_{EMNZ-2} =$ 0.22). As shown in Fig. 1(b), the high-impedance configuration, i.e. EMNZ-1 case, which exhibits the lower group index, lowers the lasing threshold, whereas the low-impedance configuration, i.e. EMNZ-2 case, increases it (see Table I). This demonstrates that a more slowed down light does not necessarily enhance light–matter interactions, consistently with the above-mentioned impedance control of light–matter interaction, i.e. the ENZ/MNZ cases.

The threshold trends observed in Fig. 1 are further confirmed by studying the carrier-density and photon density dynamics as shown in Fig. 2. Since the threshold injection current scales with the threshold carrier density $N_{th}$, the latter provides a direct measure of the threshold behavior. From Eqs. (2), the threshold carrier density is given by

$$N_{th} = N_0 + \frac{1}{Z}\frac{g_{th}}{g_0}, \qquad (3)$$

where $g_{th}$ is the threshold gain. This relation shows that, within $\varepsilon$- and μ-dispersive media, the threshold carrier density scales as the inverse of the wave impedance $Z$ (see Appendix C for the steady-state analysis of Eqs (2)).

TABLE I. Summary of the electromagnetic parameters and the corresponding lasing $I_{th}$, carrier density $N_{th}$, and photon density $S_{th}$ thresholds for the different configurations considered.

| | Ref | ENZ | MNZ | EMNZ-1 | EMNZ-2 |
|---|---|---|---|---|---|
| $\varepsilon$ | 11.56 | 0.01 | 1.00 | 0.01 | 0.20 |
| $\mu$ | 1.00 | 1.00 | 0.01 | 0.10 | 0.01 |
| $n_g$ | 4.00 | 22.01 | 22.01 | 7.61 | 10.26 |
| Z | 0.29 | 10.00 | 0.10 | 3.16 | 0.22 |
| $I_{th}$ (mA) | 15.52 | 7.52 | 31.52 | 8.01 | 18.12 |
| $N_{th}/N_0$ | 2.13 | 1.03 | 4.33 | 1.10 | 2.49 |
| $S_{th}$ $(10^{16} cm^{-3})$ | 0.83 | 5.01 | 3.71 | 1.72 | 2.06 |

Figure 2(a) shows the temporal evolution of the carrier density for the reference, ENZ, and MNZ media. The ENZ medium reaches the lasing regime at a carrier density close to $N_0$, with laser relaxation oscillations rapidly stabilizing, whereas the MNZ medium exhibits much larger carrier-density oscillations around a higher threshold. This contrasting dynamic reflects the different light–matter interactions in the two NZI classes: in ENZ medium, the enhanced electric field strengthens the interaction with the atomic polarization, allowing lasing to develop with only small variations of the population inversion. In MNZ medium, the weaker electric-field interactions require larger variations of the population inversion to produce a polarization response that is strong enough for efficient coupling to the optical field. The same behavior applies to the EMNZ class, where the carrier-density dynamics likewise depend on the impedance of the medium (Fig. 2(b)).

The field enhancement effect is further evidenced by the temporal evolution of the photon density $S$ shown in Fig. 2(c,d). The larger group index of the ENZ and MNZ media leads to an increased photon accumulation in the cavity, resulting in a higher threshold photon density $S_{\mathrm{th}}$ (Fig. 2(c)) (see Appendix C). The asymmetry between these two slow-light regimes arises from the impedance-dictated energy storage channel: the electromagnetic energy is stored predominantly in the electric field for ENZ and in the magnetic field for MNZ. For the EMNZ media (Fig. 2(d)), the group index is likewise larger than in the reference medium, leading to an increased $S_{th}$ in both cases, while the dominant energy-storage channel depends on the wave impedance. Overall, these results show that the NZI class, through its combined influence on group index and impedance, changes the carrier and photon dynamics and, consequently, the lasing threshold.

*Contact author: michael.lobet@unamur.be

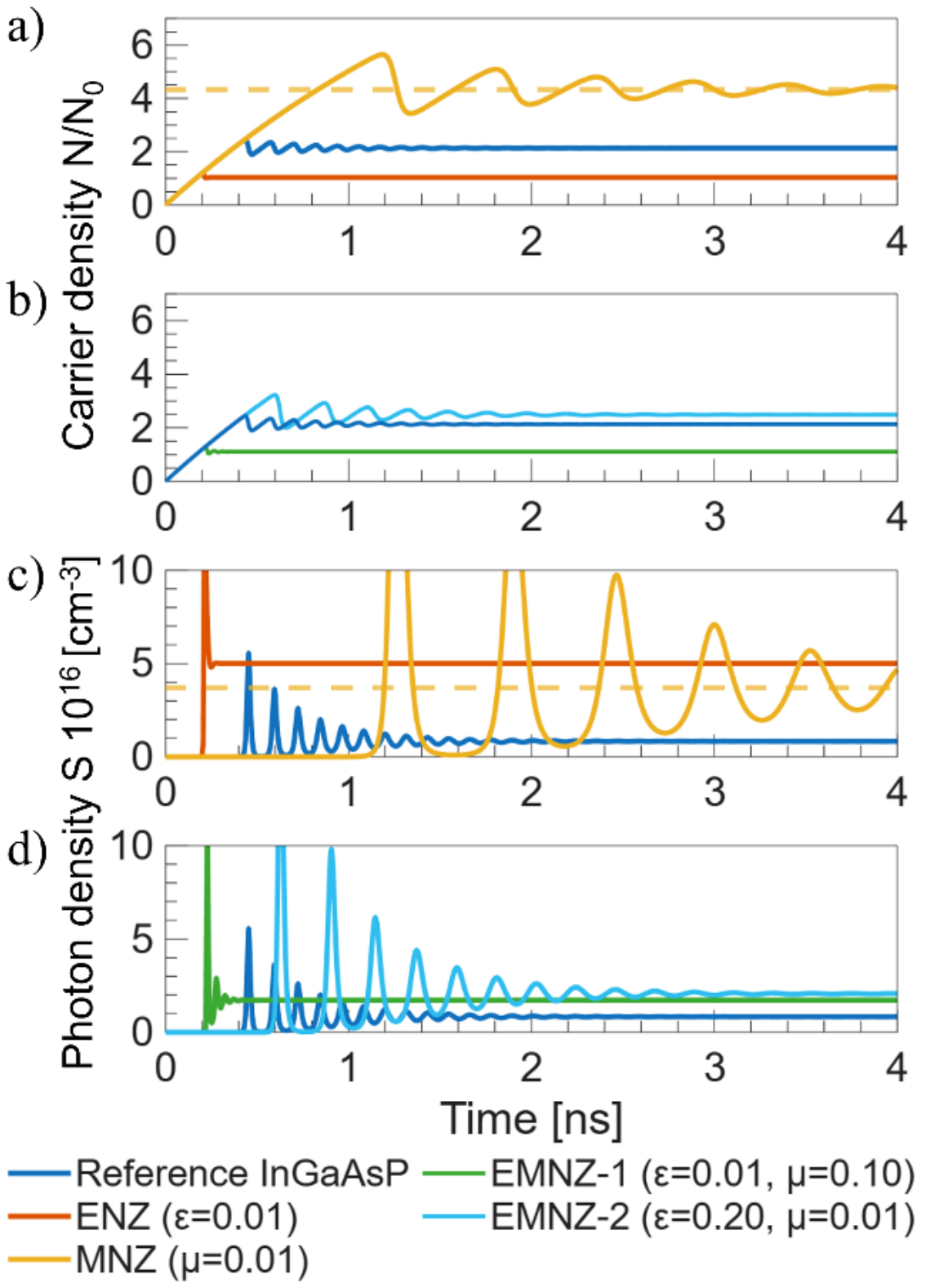

FIG 2. Temporal evolutions of (a-b) the carrier density ($N$) normalized by the transparency carrier density ($N_0$) and (c-d) the intracavity photon density ($S$) for an injection current of 100 mA. The blue, red, and yellow curves correspond to the reference InGaAsP medium, the ENZ medium, and the MNZ medium, respectively, while the light-blue and green curves correspond to the two EMNZ media. Dashed lines indicate the corresponding thresholds for MNZ.

To investigate the impact of the NZI class on the laser emission spectrum, we performed full-wave 2D-FDTD simulations with transverse-electric polarization [35]. The active medium is described by a dispersive electromagnetic response, in which both the permittivity and permeability follow frequency-dependent constitutive relations [19,35–37], with a self-consistent four-level gain model coupled to Maxwell's equations [35,38,39]. The laser cavity is modeled as a one-dimensional Fabry–Perot cavity, with the active medium enclosed between two Bragg mirrors composed of alternating InGaAsP ($n = 3.40$) and air layers satisfying the quarter-wave condition. The target wavelength $\lambda_0 = 1.3\mu$m is located at the center of the mirror's photonic band gap, and the whole cavity is embedded in air and terminated by perfectly matched layers.

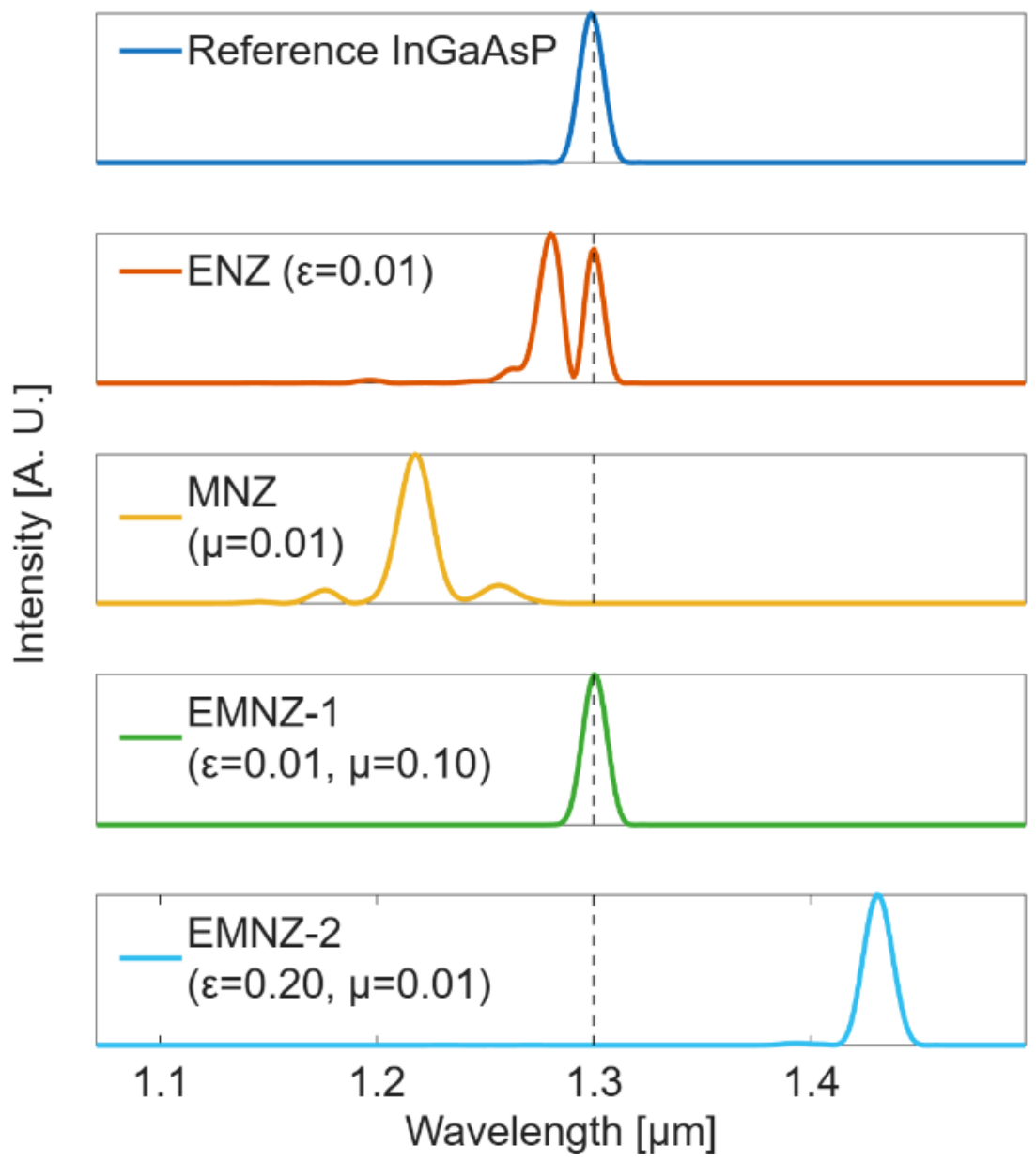

FIG 3. Emission spectra around 1.3μm (dashed line) obtained from FDTD simulations for an injection rate of $10^{16}$ s$^{-1}$. The blue, red, yellow, green and light blue curves correspond to the reference InGaAsP, ENZ, MNZ, EMNZ-1 and EMNZ-2 media, respectively. The cavity lengths are $L = 5\lambda_0/n$ for the reference and $L = \lambda_0/(2n)$ for the ENZ, MNZ and EMNZ media.

The emission spectra (Fig. 3) reveal distinct modal behaviors across the NZI classes. The reference InGaAsP cavity exhibits a single emission peak near 1.3μm, corresponding to stable single-mode operation. In contrast, the ENZ and MNZ media exhibit additional longitudinal modes around the lasing frequency, characteristics of multimode emission. The EMNZ cavity retains a dominant single emission peak. This behavior is governed by the cavity mode spacing, $\Delta\nu = c/(2n_gL)$ [30]: the large group index of ENZ and MNZ ($n_g = 22.01$) reduces the longitudinal mode spacing, allowing additional modes to emerge, whereas the moderate group index of the EMNZ preserves a sufficiently large mode spacing for single-mode operation to prevail. The frequency shift observed in the EMNZ-2 medium with the lowest impedance arises from the dispersion of $\varepsilon$ and μ. The peak occurs where the permittivity approaches zero, indicating that, in the cavity mode competition, the mode with the lowest permittivity, and hence the highest impedance, is preferentially favored.

*Contact author: michael.lobet@unamur.be

Interestingly, the single-mode emission observed for the EMNZ media and the multimode emission of the ENZ/MNZ media are consistent with previously reported experiments. Contractor *et al.* demonstrated that a photonic-crystal laser featuring a Dirac cone at the Γ-point maintained single-mode operation over large cavity areas, whereas lifting the Dirac-cone degeneracy led to multimode emission [40]. Such Dirac-cone photonic crystals have been shown to exhibit an effective EMNZ response, while lifting the degeneracy drives the system toward ENZ or MNZ responses [12,41,42]. Our results provide a unified interpretation of these observations through the group index dependence of the NZI classes.

In conclusion, we have theoretically demonstrated that each of the three classes of NZI media provides its own direct control over laser emission properties. By developing laser rate equations that explicitly incorporate the electric and magnetic dispersive properties of the active medium, we revealed that the impedance is the key parameter governing the lasing threshold. The high impedance of ENZ media lowers the lasing threshold through enhanced electric-field interactions, whereas the low impedance of MNZ media makes them less favorable for low-threshold lasing due to their weaker electric-field-mediated light–matter interactions. In EMNZ media, the tunable impedance provides an additional degree of control over the lasing threshold. Numerical simulations further showed that the slow-light response of ENZ and MNZ media promotes the emergence of additional spectral modes due to a reduced cavity-mode spacing, whereas the lower group index of EMNZ media helps preserve single-mode emission. These results establish a direct link between the electromagnetic properties of NZI media and laser operation, opening new opportunities for the design of photonic-crystal lasers [40], where Dirac-cone dispersions, for instance, provide a natural route toward effective NZI responses [42]. Furthermore, the extension of this approach to time-varying metamaterials could enable dynamic control of lasing properties through actively tunable electromagnetic responses [43–45].

## ACKNOWLEDGMENTS

The authors would like to thank J. Faist for stimulating discussion. A. D. is a research fellow of the Fonds de la Recherche Scientifique – FNRS. M.L. is a Research Associate of the Fonds de la Recherche Scientifique – FNRS. This research used resources of the "Plateforme Technologique de Calcul Intensif (PTCI)" (http://www.ptci.unamur.be) located at the University of Namur, Belgium, which is supported by the FNRS-FRFC, the Walloon Region, and the University of Namur (Conventions No. 2.5020.11, GEQ U.G006.15, 1610468, RW/GEQ2016 et U.G011.22). The PTCI is member of the "Consortium des Équipements de Calcul Intensif (CÉCI)" (http://www.ceci-hpc.be). Generative AI tools were used during the preparation of this manuscript to assist with text drafting, language editing, and for translation purposes only. All AI-generated content was subsequently reviewed, modified, and validated by the authors, who take full responsibility for the final content of the manuscript.

## APPENDIX A: Maxwell wave equation in a dispersive cavity

We derive hereafter the evolution equation of the electric field $\boldsymbol{\mathcal{E}}$ inside a laser cavity containing a dispersive medium. We start from the wave equation in a dispersive medium [46],

$$\nabla^2\boldsymbol{\mathcal{E}}(\boldsymbol{r},t) - \frac{n^2(\omega)}{c^2}\frac{\partial^2\boldsymbol{\mathcal{E}}(\boldsymbol{r},t)}{\partial t^2} = \frac{\mu(\omega)}{\varepsilon_0 c^2}\frac{\partial^2\boldsymbol{P_{at}}(\boldsymbol{r},t)}{\partial t^2}, \quad (4)$$

where $\omega$ is the angular frequency, $c$ the speed of light and $\varepsilon_0$ the vacuum permittivity. The phase index $n(\omega)$ and permeability $\mu(\omega)$ account for the dispersive properties of the medium and $\boldsymbol{P}_{at}$ is the atomic polarization. Inside the cavity, the electric field is expanded over the cavity eigenmodes as

$$\boldsymbol{\mathcal{E}}(\boldsymbol{r},t) = e^{-i\omega_0 t}\sum_q E_q(t)e^{i\boldsymbol{k_{p,q}}\cdot\boldsymbol{r}}\boldsymbol{u_q}(\boldsymbol{r}), \quad (5)$$

where $E_q(t)$, $\boldsymbol{k_{p,q}}$ and $\boldsymbol{u_q}(\boldsymbol{r})$ denotes the slowly varying field envelope, the complex wavevector and the spatial distribution of mode $q$, respectively [32]. Each cavity mode induces a corresponding polarization $P_q(t)$, allowing the atomic polarization $\boldsymbol{P_{at}}$ to be expanded on the same modal basis [32]. Substituting these modal expansions into the wave equation and working in the frequency domain yields an evolution equation for the modal amplitudes $E_q(\omega)$.

Thanks to the slowly varying envelope approximation [30,32], spatial derivatives of the slowly varying envelopes can be neglected with respect to those of the rapidly varying phase factors, such that

$$\nabla^2\left[e^{i\boldsymbol{k_{p,q}}\cdot\boldsymbol{r}}\boldsymbol{u_q}(\boldsymbol{r})\right] \simeq -k_{p,q}^2 e^{i\boldsymbol{k_{p,q}}\cdot\boldsymbol{r}}\boldsymbol{u_q}(\boldsymbol{r}). \quad (6)$$

The propagation wavevector $k_p$ is first expanded to second order around the central frequency $\omega_0$, then squared and only the lowest order derivatives are retained:

*Contact author: michael.lobet@unamur.be

$$k_p^2(\omega) \simeq k_{p,0}^2 + \frac{2}{v_g} k_{p,0}(\omega - \omega_0) + \frac{1}{2} k_{p,0}\sigma(\omega - \omega_0)^2 \quad (7)$$

where $k_{p,0} = k_p(\omega_0)$, $v_g = \frac{dk_p}{d\omega}|_{\omega_0}$ is the group velocity and $\sigma = \frac{d^2k_p}{d\omega^2}|_{\omega_0}$is the group-velocity-dispersion coefficient. Assuming that the polarization envelope varies also slowly in time, the material response is evaluated at the central frequency by the approximate expression:

$$\mu(\omega)\omega^2 P_q(\omega - \omega_0) \simeq \mu(\omega_0)\omega_0^2 P_q(\omega - \omega_0). \quad (8)$$

After performing inverse Fourier transformation, the modal amplitudes satisfy

$$\sum_q \left[ (k_{p,0}^2 - k_{p,q}^2) E_q + \frac{2ik_{p,0}}{vg}\frac{dE_q}{dt} + k_{p,0}\sigma \frac{d^2E_q}{dt^2} \right] e^{i\boldsymbol{k}_{p,q}\cdot\boldsymbol{r}} \boldsymbol{u_q}(\boldsymbol{r}) = -\frac{\mu(\omega_0)\omega_0^2}{\varepsilon_0 c^2} \sum_q P_q e^{i\boldsymbol{k}_{p,q}\cdot\boldsymbol{r}} \boldsymbol{u_q}(\boldsymbol{r}). \quad (9)$$

We now assume that the field frequency is close to a single cavity resonance and that mode spacing is sufficiently large to neglect all other modes [32]. The modal expansion therefore reduces to a single mode characterized by the complex wavevector $\tilde{k}_{p,c}$. The field equation (9) becomes

$$\frac{1}{v_g}\frac{dE}{dt} - \frac{i(k_{p,0}^2 - \tilde{k}_{p,c}^2)}{2k_{p,0}} E = \frac{i\mu(\omega_0)\omega_0}{2\varepsilon_0 n(\omega_0)c} P_{at}, \quad (10)$$

where we additionally consider a negligible group-velocity-dispersion coefficient ($\sigma \simeq 0$). Cavity losses are introduced through the imaginary part of the resonance wavevector, $\tilde{k}_{p,c} = k_{p,c} - \frac{i\alpha_c}{2}$, where $\alpha_c$ is the total cavity loss coefficient. Defining the cavity decay rate ($\gamma_c = v_g\alpha_c$), we finally obtain

$$\frac{dE}{dt} = iv_g(k_{p,0} - k_{p,c})E - \frac{\gamma c}{2} E + v_g \frac{i\mu(\omega_0)\omega_0}{2\varepsilon_0 n(\omega_0)c} P_{at}. \quad (11)$$

which corresponds to Eq. (1) of the main text with $Z(\omega_0) = \mu(\omega_0)/n(\omega_0)$.

## APPENDIX B: From Maxwell-Bloch equations to laser rate equations

We briefly show how the semiconductor laser rate equations can be deduced from the Maxwell–Bloch equations. In a semiconductor, the microscopic quantities, i.e., electrons, possess de Broglie wavevector $\boldsymbol{k_e}$ [30]. The microscopic polarization $p_{\boldsymbol{k_e}}$, the electronic transition frequency $\omega_{a,\boldsymbol{k_e}}$, the detuning $\delta_{\boldsymbol{k_e}} = \omega_0 - \omega_{a,\boldsymbol{k_e}}$, the dipole moment $d_{\boldsymbol{k_e}}$, and the carrier populations in the conduction and valence bands therefore become dependent on $\boldsymbol{k_e}$. The population inversion $w_{\boldsymbol{k_e}}$ is written as

$$w_{\boldsymbol{k_e}} = N_{e,\boldsymbol{k_e}} + N_{h,\boldsymbol{k_e}} - 1, \quad (12)$$

where $N_{e,\boldsymbol{k_e}}$ and $N_{h,\boldsymbol{k_e}}$ denote the occupation probabilities of electrons and holes, respectively. The semiconductor laser Bloch equations then read [30]

$$\frac{dp_{\boldsymbol{k_e}}}{dt} = -i\delta_{\boldsymbol{k_e}} p_{\boldsymbol{k_e}} - \frac{i}{2}\Omega_{\boldsymbol{k_e}}(N_{e,\boldsymbol{k_e}} + N_{h,\boldsymbol{k_e}} - 1) - \gamma_\perp p_{\boldsymbol{k_e}},$$
$$\frac{dN_{l,\boldsymbol{k_e}}}{dt} = \frac{i}{2}(\Omega_{\boldsymbol{k_e}} p_{\boldsymbol{k_e}}^* - \Omega_{\boldsymbol{k_e}}^* p_{\boldsymbol{k_e}}) + \Lambda_{l,\boldsymbol{k_e}} - \gamma_{nr} N_{l,\boldsymbol{k_e}} - \gamma_{eq}(N_{l,\boldsymbol{k_e}} - f_{F,l,\boldsymbol{k_e}}), \quad (13)$$

where $\Omega_{\boldsymbol{k_e}} = d_{\boldsymbol{k_e}} E/\hbar$ is the Rabi frequency, $l$ denotes electrons ($e$) or holes ($h$), $\Lambda_{l,\boldsymbol{k_e}}$ describes carrier injection, $\gamma_{nr}$ is the non-radiative recombination rate, $f_{F,l,\boldsymbol{k_e}}$ denotes the carrier quasi-equilibrium Fermi distribution, and $\gamma_\perp$ is the polarization decay rate. Since $\gamma_\perp$ is typically much larger than both the carrier and field evolution rates [30,32], the microscopic polarization can be adiabatically eliminated by setting $\frac{dp_{\boldsymbol{k_e}}}{dt} = 0$. This approximation yields,

$$p_{\boldsymbol{k_e}} = -\frac{i}{2}\frac{N_{e,\boldsymbol{k_e}} + N_{h,\boldsymbol{k_e}} - 1}{\gamma_\perp + i\delta_{\boldsymbol{k_e}}}\Omega_{\boldsymbol{k_e}}. \quad (14)$$

The carrier density $N$ and the macroscopic polarization $P_{at}$ are respectively defined as [30],

$$N = \frac{1}{V}\sum_{\boldsymbol{k_e}} N_{e,\boldsymbol{k_e}} = \frac{1}{V}\sum_{\boldsymbol{k_e}} N_{h,\boldsymbol{k_e}} \quad (15)$$

and

$$P_{at} = \frac{2}{V}\sum_{\boldsymbol{k_e}} d_{\boldsymbol{k_e}} p_{\boldsymbol{k_e}}, \quad (16)$$

where $V$ is the lasing medium volume. Under the quasi-equilibrium approximation, the carrier distributions remain close to their Fermi–Dirac distributions, i.e., $N_{l,\boldsymbol{k_e}} \cong f_{F,l,\boldsymbol{k_e}}$, allowing the resulting equations to be expressed solely in terms of the carrier density $N$ [30].

Substituting the adiabatic polarization (14) into the macroscopic polarization (16) leads naturally to the definition of a complex gain [30,47],

$$\tilde{g}(N) = \frac{\omega_0}{c\varepsilon_0 V\hbar}\sum_{\boldsymbol{k_e}} d_{\boldsymbol{k_e}}^2 \frac{(\gamma_\perp - i\delta_{\boldsymbol{k_e}})}{\gamma_\perp^2 + \delta_{\boldsymbol{k_e}}^2}(f_{F,e,\boldsymbol{k_e}} + f_{F,h,\boldsymbol{k_e}} - 1), \quad (17)$$

*Contact author: michael.lobet@unamur.be

whose real part corresponds to the material gain and imaginary part accounts for carrier-induced refractive-index variations. The Maxwell–Bloch equations therefore reduce to

$$\frac{dE}{dt} = iv_g\left(k_{p,0} - k_{p,c}\right)E - \frac{\gamma_c}{2}E + \frac{1}{2}Zv_g\tilde{g}(N)E,$$

$$\frac{dN}{dt} = -v_g Z g(N)\frac{n_g\varepsilon_0}{2Z\hbar\omega_0}|E|^2 + \frac{J}{eV} - \gamma_{nr}N, \qquad (18)$$

where $g(N) = Re[\tilde{g}(N)]$. Finally, defining the photon density as $S = \frac{n_g\varepsilon_0}{2Z\hbar\omega_0}|E|^2$ [46], one recovers the laser rate equations given in Eq. (2) of the main text.

## APPENDIX C: Steady-state analysis

The steady-state analysis of Eqs 2 allows us to analytically evaluate the expression for the photon density threshold in the cavity:

$$S_{th} = \frac{1}{v_g g_{th}}\left(\Lambda_P - \gamma_{nr}N_{th}\right), \qquad (19)$$

showing that $S_{th}$ is primarily affected by group velocity $v_g$ and, to a lesser extent, by impedance Z through $N_{th}$, as expressed by Eq. (3). Furthermore, as expected, a lower threshold leads to a shorter time, $t_d$, to reach lasing threshold.

$$t_d = \frac{1}{\gamma_{nr}}\ln\left(\frac{\Lambda_p}{\Lambda_p - \gamma_{nr}N_{th}}\right). \qquad (20)$$

Finally, the laser oscillation period $\omega_r$ is affected by the impedance and the group velocity,

$$\omega_r^2 = \frac{Z}{v_g}g_0\left(\Lambda_p - \gamma_{nr}N_{th}\right), \qquad (21)$$

while the damping constant $\gamma_0$ is affected only by the impedance:

$$\gamma_0 = Z\frac{g_0}{g_{th}}\left(\Lambda_p - \gamma_{nr}N_{th}\right). \qquad (22)$$

*Contact author: michael.lobet@unamur.be

*Contact author: michael.lobet@unamur.be